\documentclass[fleqn,12pt,twoside]{article}
\usepackage{espcrc1}
\usepackage{graphicx}
\usepackage[figuresright]{rotating}

\newcommand{\AmS}{{\protect\the\textfont2
  A\kern-.1667em\lower.5ex\hbox{M}\kern-.125emS}}

\title{$\Omega^-$ and ${\overline{\Omega}}^{+}$ production in Au+Au collisions at $\sqrt{s_{NN}}=130$~and~$200$~GeV}

\author{C. Suire \address{Brookhaven National Laboratory,\\ 
   		  Upton, NY 11973}
	for the STAR Collaboration \thanks{For the full author list and acknowledgements, see Appendix "Collaborations" of this volume} 
	}

\begin{document}

\maketitle

\begin{abstract}
Mid-rapidity Omega and anti-Omega production in Au+Au collisions at RHIC is studied with the STAR experiment. We report preliminary results on yields and spectra at  $\sqrt{s_{NN}}=130$ and $200$~GeV. Production relative to negatively charged hadrons (h$^-$) as well as thermal freeze-out and collective expansion are discussed.
\end{abstract}

\section{Introduction}
Strangeness production/enhancement has been long suggested as a possible signature able to discriminate  between the quark-gluon plasma (QGP) and hadron gas (HG) scenario in heavy ions collisions.  Due to high gluon density expected in high energy nuclear collisions, the production of multi-strange baryons ($\Xi$, $\Omega$) may be enhanced over other particles \cite{Koch86}. 
Furthermore, any enhancement is expected to increase with the strangeness content of the particle, making the $\Omega$ the most sensitive strangeness probe. Its production relative to other hadrons gives a severe constraint to statistical models.

In addition, early kinetic freeze-out of $\Omega$, $J/\Psi$ and $\Psi^{\prime}$ has been suggested at SPS, which would result in these particles carrying information on the flow velocity of strongly interacting matter (HG or QGP at hadronisation) \cite{Buga02}. Thus, measuring the transverse velocity of $\Omega$ may be  helpful in understanding the system evolution of collective expansion~\cite{Tean01}.

\section{Experiment and Analysis}
 The Solenoidal Tracker at RHIC (STAR) detector system \cite{Harr01}, in the configuration used to collect data presented here, consists principally of a large Time Projection Chamber (TPC) used for charged particle tracking. A magnet surrounding the TPC was operated at 0.25 (0.5) T for the data taken at $\sqrt{s_{NN}} = 130 \,\, (200)$~GeV. Triggering was achieved using two hadronic calorimeters (ZDC) in the very forward region and an array of scintillator slats arranged in  a barrel (CTB) around the TPC. The centrality  cuts used correspond to the top 11\% (130 GeV) and 10\% (200 GeV) of  the total hadronic cross section. In addition, the reconstructed $\Omega$ were selected in the kinematical region $p_{\perp} > 0.5$ GeV/c and $|y| < 0.75$. The total number of events analysed is $0.415 \times  10^6$  and $1.38 \times 10^6$ for the 130 and 200 GeV runs respectively. 

\begin{figure}[h!]
  \begin{minipage}[h]{.48\linewidth} 
  \includegraphics[width=8.0cm,height=5.5cm,clip=true,draft=false]
  	{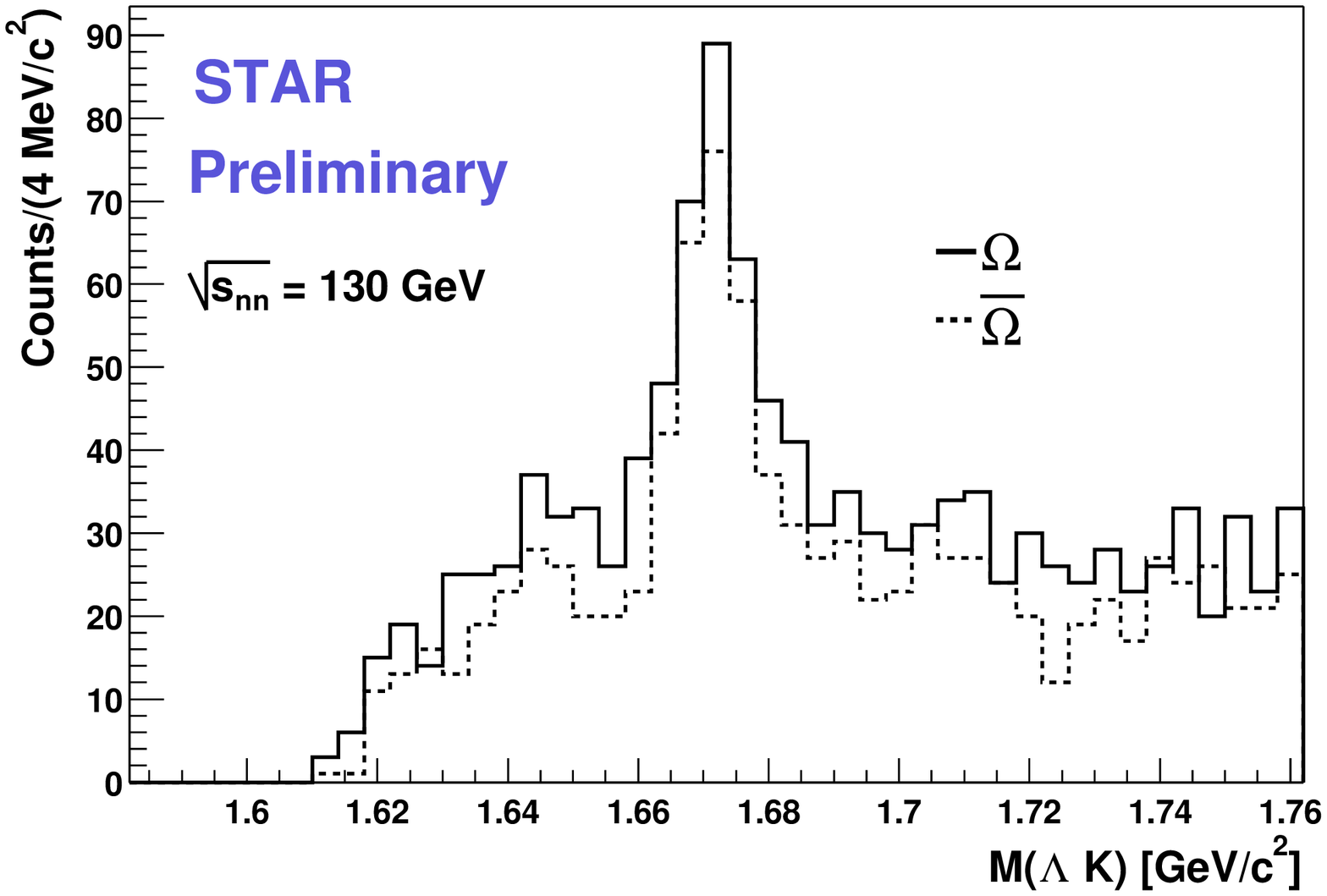}
  \end{minipage} \hfill
  \begin{minipage}[h]{.48\linewidth}
  \includegraphics[width=8.0cm,height=5.5cm,clip=true,draft=false]
   	{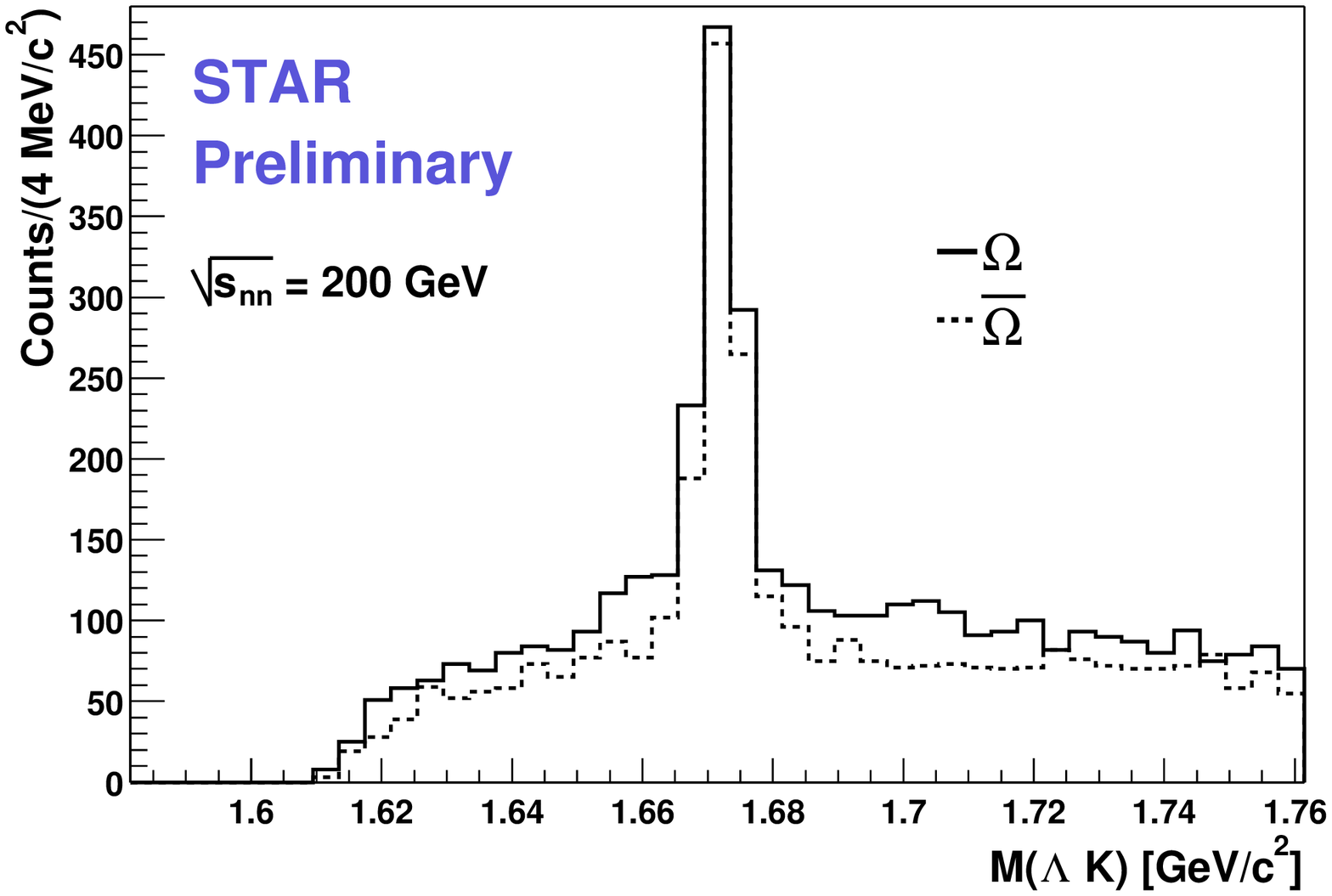}
  \end{minipage}
  \vspace{-0.5cm}
  \caption{Invariant mass distribution for ($\Lambda K^-$)  and (${\overline{\Lambda}} K^+$) at two energies.}
  \label{fig:massInv}
\end{figure}

The current analysis proceeds as follows in order to reconstruct the Omega baryon. As a first step, $\Omega^-$ were  reconstructed topologically through the sequential weak decay  $\Omega \rightarrow K^- + \Lambda$ (BR=67.8\%) and $\Lambda  \rightarrow p + \pi^-$(BR=63.9\%).
Within an event, all tracks reconstructed in the TPC  were  extrapolated backward to the primary vertex of the collision. Candidate tracks were paired to form  neutral decay vertices ($V0$), which were  associated afterward with a bachelor track to form $\Omega$ vertices. Selections are mainly based on geometrical cuts, i.e. distance of closest approach between candidate tracks and to the primary vertex.
 Then, to reduce further the combinatorial  background resulting from close but uncorrelated tracks, an identification  of daughter tracks based on mean specific energy loss in the TPC  was used. Invariant mass distributions of the $\Omega^-$ and ${\overline{\Omega}}^+$ are shown on Fig.\ref{fig:massInv}. 
\vspace{-0.2cm}
\vspace{-0.2cm}
To estimate the raw signal, the background was subtracted by using a linear interpolation between two regions located a few MeV/c$^2$ on either side of the peak region. The raw signals were  corrected for acceptance and efficiency using Monte Carlo $\Omega$'s embedded into real events.

\section{Results and Discussion}
\begin{figure}[h!]
  \begin{minipage}[h]{.6\linewidth} 
  \includegraphics[scale=0.5,draft=false]
  	{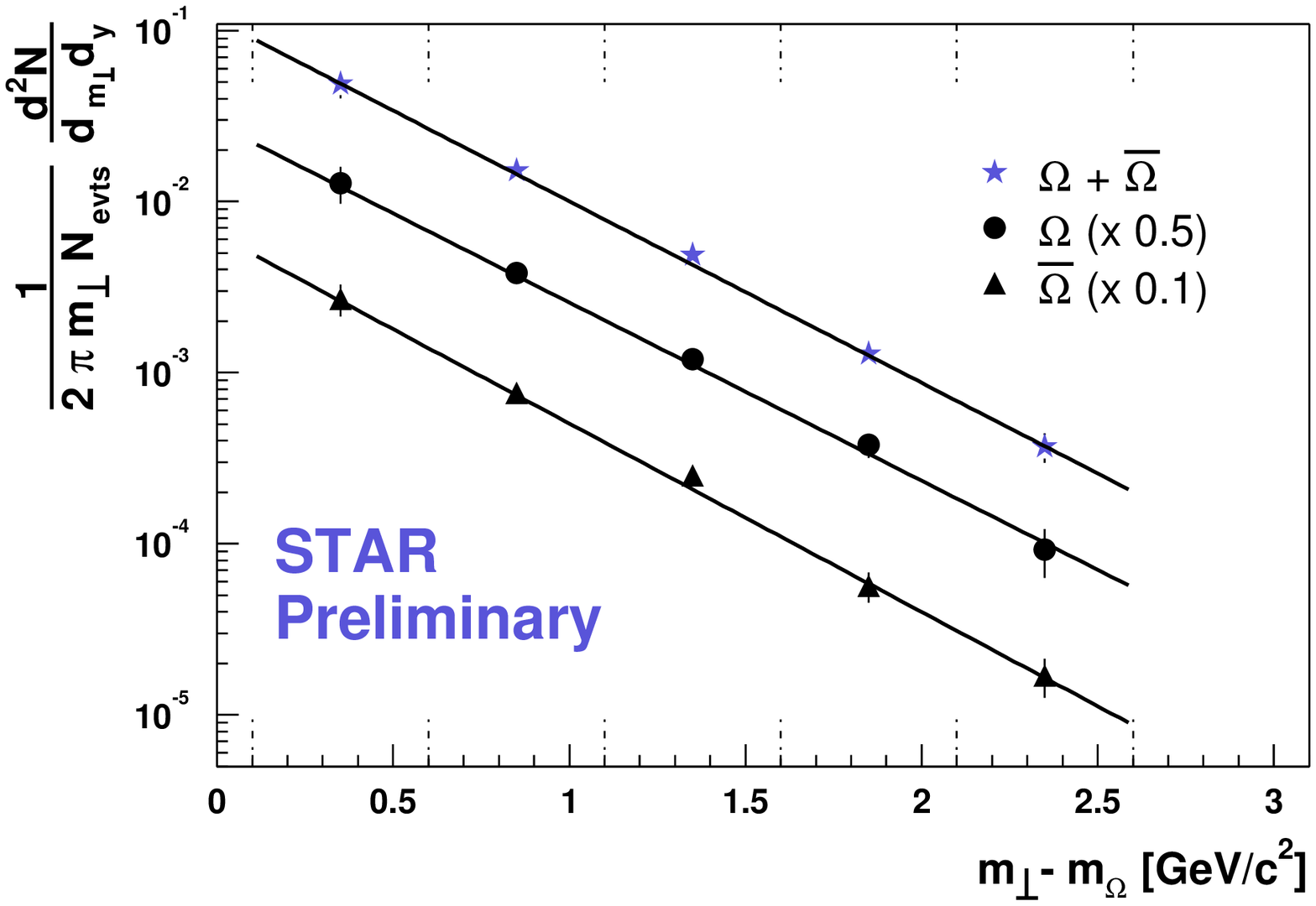}
  \end{minipage} \hfill
  \begin{minipage}[h]{.4\linewidth}
  \includegraphics[scale=0.32,draft=false]
   	{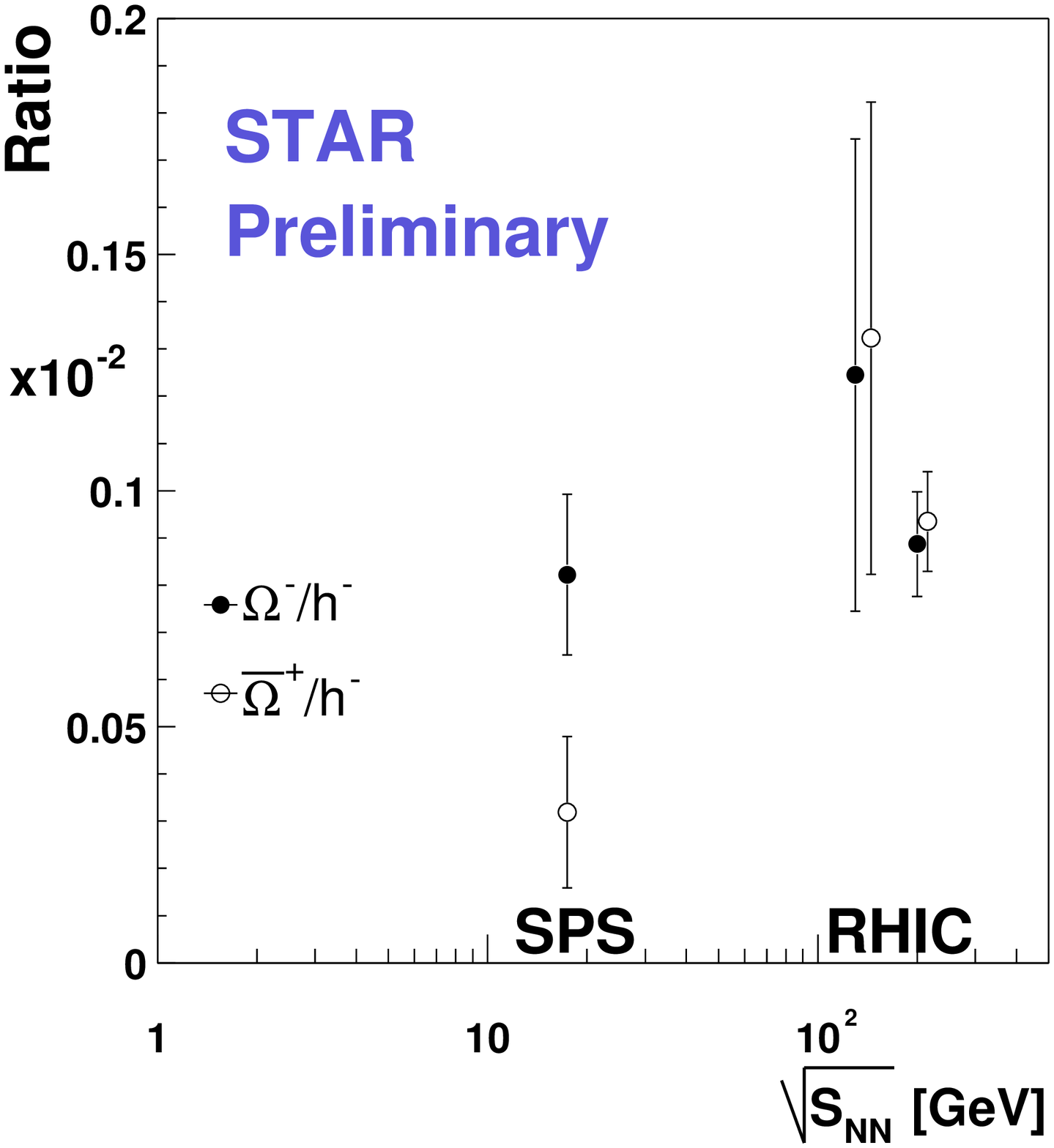}
  \end{minipage}
  \vspace{-0.7cm}
  \caption{On the left part, corrected transverse mass spectra for  $\Omega^-$ + ${\overline{\Omega}}^+$, $\Omega^-$ and $\overline{\Omega}^+$ at $\sqrt{s_{NN}}~=~200$~GeV (dotted vertical bars indicate bin size). On the right part, $\Omega / h^-$ as a function of beam energy (stat. errors only). SPS values are from \cite{Cali99}.}
  \label{fig:ratioOmHminus}
\end{figure}

Figure \ref{fig:ratioOmHminus} shows the corrected transverse mass spectra at $\sqrt{s_{NN}} = 200 $ GeV. The yields (dN/dy) and the inverse slopes (T) are  extracted from exponential fits (specified by eq.\ref{eq:fitFunction}) to the data. 
\begin{equation}
\frac{1}{2 \pi m_{\perp}} \frac{d^2N}{dm_{\perp} dy} = \frac{dN/dy}{2 \pi T (m_{\Omega} + T)}  \exp  \left\{ -  \frac{m_{\perp} - m_{\Omega}}{T} \right\} \quad \mbox{where} \quad m_{\perp}=\sqrt{p_{\perp}^2 + m_{\Omega}^2}
\label{eq:fitFunction}
\end{equation}
The results are quoted in table~\ref{table:yieldsSlopes} along with those  at $\sqrt{s_{NN}}$ = 130 GeV \cite{Hipp02}. Statistical models have produced values of dN/dy that are in  good agreement with preliminary results at $\sqrt{s_{NN}} = 130 $ GeV \cite{Munz01,Gene02,Bron02}, assuming chemical equilibrium with T $\simeq$ 175 MeV and $\mu_B$  $\simeq$ 47 MeV. 
\begin{table}[h!]
\vspace{-0.2cm}
\caption{Yields and inverse slopes parameters (extrapolations at low m$_{\perp}$ correspond to 34\% and 18\% of the total yields). Errors quoted are statistical only, systematic errors are estimated around 15\%.}
\label{table:yieldsSlopes}
\newcommand{\m}{\hphantom{$-$}}
\newcommand{\cc}[1]{\multicolumn{1}{c}{#1}}
\renewcommand{\arraystretch}{1.2} 
\begin{tabular}{@{}l|ll|ll}
\hline
Energy          & \multicolumn{2}{c|}{$\sqrt{s_{NN}} = 130$ GeV } &  \multicolumn{2}{c}{$\sqrt{s_{NN}} = 200$ GeV}    \\
\hline
Particle        & $\Omega^-$ & $\overline{\Omega}^+$  & $\Omega^-$ & $\overline{\Omega}^+$   \\
T (MeV)         & 424  $\pm$ 65  & 387  $\pm$ 52   & 417  $\pm$ 23  & 394  $\pm$ 19  \\
dN/dy           & 0.32 $\pm$ 0.09& 0.34 $\pm$ 0.09 & 0.31 $\pm$ 0.04& 0.33 $\pm$ 0.05 \\
\hline
\end{tabular}
\vspace{-0.2cm}
\end{table}
The mid-rapidity ratios $\Omega^- / h^-$ and $\overline{\Omega}^+ / h^-$  are shown on Fig.~\ref{fig:ratioOmHminus} for SPS and RHIC energies. Unlike at the SPS, the $\overline{\Omega}^+ / h^-$ ratio approaches the $\Omega^- / h^-$ ratio at RHIC; this can be explained by the lower  net baryon density in the system.  The $\Omega^- / h^-$ ratio is found to be consistent  with the values obtained at SPS; its value might indicate that strangeness saturation was already reached  at SPS and we are still in that regime at RHIC. This trend has also been observed for the $\Xi$ particles and is detailed in~\cite{Jecc02}.
\vspace{-0.2cm}
\vspace{-0.2cm}

\begin{figure}[h!]
  \begin{minipage}[h]{.6\linewidth} 
  \includegraphics[scale=0.5,draft=false]
	{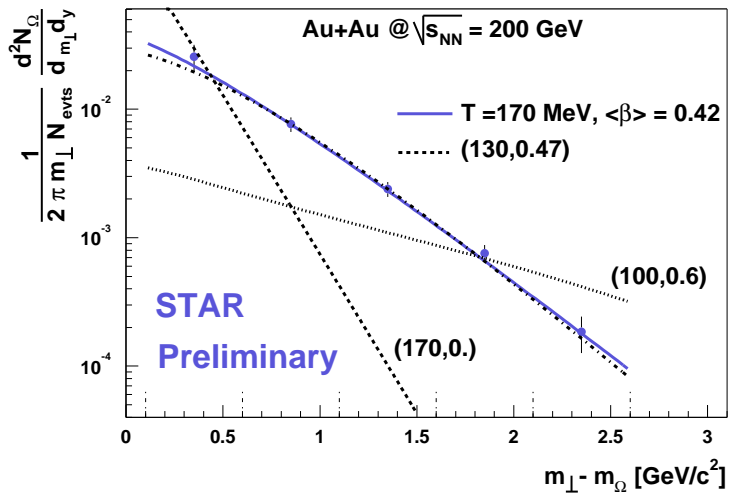}
  \end{minipage} \hfill
  \begin{minipage}[h]{.35\linewidth}
	\label{fig:omegaBWFit}
	\vspace{-1.8cm}
	\caption{Using  different blast-wave  parameterization  for $\Omega^-$. Kinetic freeze-out temperature is not a free parameter (T$_{fo}^{ki}$  =  170 MeV or 130 MeV). The two other lines are a representation of the fit function for two extreme cases : (T$_{fo}^{ki}$ = 100 MeV, $\langle\beta_{T}\rangle$ = 0.6) 
and (T$_{fo}^{ki}$ = 170 MeV, $\langle\beta_{T}\rangle$ = 0.)}
  \end{minipage}
\end{figure}

To address the question of transverse flow, m$_{\perp}$ spectra were fit using the  so-called blast-wave parameterization \cite{Schn93} :
\begin{equation}
\frac{dN}{ m_{\perp} dm_{\perp}} \propto \int_0^R rdr \, m_{\perp} I_0  (\frac{p_{\perp} \sinh \rho}{T}) K_1(\frac{m_{\perp} \cosh \rho}{T}) \quad \mbox{with} \quad
\begin{array}{ll}
&\rho  = \tanh^{-1} \beta_{T}(r),\\
&\beta_{T}(r) = 1.4 \langle\beta_{T}\rangle (r/R)^{0.8}. 
\end{array}
\label{eq:blastWawe}
\end{equation}
To perform the first fit (Fig.\ref{fig:omegaBWFit}) the kinetic freeze-out temperature has been set to the value of chemical freeze-out  given by statistical models (T$_{fo}^{ki}$ = T$_{fo}^{ch}$ $\sim$ 170 MeV). Thus, we obtain $\langle\beta_{T}\rangle$ equal to 0.45 $\pm$ 0.17 (syst.). This compares to $\langle\beta_{T}\rangle_{SPS}$ $\simeq$ 0.2  for $\Omega, J/\Psi, \Psi^{\prime}$ \cite{Buga02} and may be interpreted as a higher transverse flow component at RHIC. The parameters extracted from similar fits for pions, kaons and protons are T$_{fo}^{ki}$  = 100 MeV and $\langle\beta_{T}\rangle$ = 0.6 (for Au+Au at  200 GeV) \cite{Reti02}, and are not compatible with the ones we obtained for $\Omega$ (using a different approach, the latest NA49 results \cite{Leuv02} have shown that a global fit of particles spectra with a single set of parameters (T$_{fo}^{ki}$,$\langle\beta_{T}\rangle$), including $\Omega$, was possible  at $\sqrt{s_{NN}}$ = 17.3 GeV).  
Nevertheless, the two parameters (T$_{fo}^{ki}$,$\langle\beta_{T}\rangle$) are strongly correlated and others values can be used that give equally good results, as demonstrated with the second fit on Fig.~\ref{fig:omegaBWFit} fixing T$_{fo}^{ki}$ at 130 MeV. Further  studies may  allow us  to determine (T$_{fo}^{ki}$,$\langle\beta_{T}\rangle$) more precisely  and hence  estimate the limiting values for these parameters. 
\vspace{-0.4cm}
\vspace{-0.2cm}
\vspace{-1.0cm}
\section{Conclusion}
We have presented preliminary results on $\Omega$ and ${\overline{\Omega}}^{+}$ production at RHIC.  We found similar $dN/dy$ and inverse slope values at $\sqrt{s_{NN}}$ = 130 GeV and 200 GeV. Multi-strange particles ratios  give some indication that the fireball is chemically equilibrated and the $\Omega$ production scale with h$^-$ from SPS to RHIC. The mean transverse velocity of the $\Omega$, assuming an early thermal freeze-out, is found to be $\langle\beta_{T}\rangle$ = 0.45 $\pm$ 0.17 (syst.). It would be premature to draw conclusions about $\Omega$ flow (since, among other details, its $\sigma_{elast}$ is unknown) but the increase in $\langle\beta_{T}\rangle$  from SPS to RHIC energy raises the question of which mechanism might  produce it \cite{Tean01}. In the near future, with larger statistics, improvements in the determination of transverse flow (blast-wave fits) as well as  anisotropic flow (v$_{2}$) measurement of $\Omega$ might be possible and give us a clearer picture of the $\Omega$ dynamics.

\end{document}